\documentstyle[12pt,epsfig]{article}
\newcommand{\beq}{\begin{equation}}
\newcommand{\eeq}{\end{equation}}
\newcommand{\beqa}{\begin{eqnarray}}
\newcommand{\eeqa}{\end{eqnarray}}

\newcommand{\vs}{\vspace{-0.25cm}}

\begin{document}

\hfill FZJ-IKP(TH)-1999-04


\vspace{1in}

\begin{center}

{{\large\bf Charge independence breaking and charge symmetry breaking \\[0.3em]
in the nucleon--nucleon interaction from effective field theory}}

\end{center}

\vspace{.3in}
\begin{center} 
E.~Epelbaum,$^{a,b}$\footnote{email: evgeni.epelbaum@hadron.tp2.ruhr-uni-bochum.de}  
Ulf-G. Mei{\ss}ner$^b$\footnote{email: Ulf-G.Meissner@fz-juelich.de}

\bigskip

$^a${\it Ruhr-Universit\"at Bochum, Institut f{\"u}r
  Theoretische Physik II\\ D-44870 Bochum, Germany} 

\bigskip

$^b${\it Forschungszentrum J\"ulich, Institut f\"ur Kernphysik 
(Theorie)\\ D-52425 J\"ulich, Germany}

\end{center}

\vspace{.9in}
\thispagestyle{empty} 

\begin{abstract}\noindent
We discuss charge symmetry and charge independence breaking in an effective
field theory approach for few--nucleon systems. We systematically introduce
strong isospin--violating and electromagnetic operators in the theory. The
charge dependence observed in the nucleon--nucleon scattering lengths
is due to one--pion
exchange and one electromagnetic four--nucleon contact term. This
gives a parameter free expression for the
charge dependence of the corresponding effective ranges, which is
in agreement with the rather small and uncertain empirical
determinations.
We also compare the low energy phase shifts of the $nn$ and
the $np$ system.
\end{abstract}

\vspace{1.5cm}

\begin{center}
{\small\textsc DEDICATED TO WALTER GL\"OCKLE ON THE OCCASION OF HIS $60^{\rm th}$
BIRTHDAY}
\end{center}

\vfill

\pagebreak


\noindent {\bf 1.} It is well established that the nucleon--nucleon interactions
are charge dependent (for a review, see e.g.\cite{gerry}). For example, in 
the $^1S_0$ channel one has for the scattering lengths $a$ and the effective
ranges $r$ ($n$ and $p$ refers to the neutron and the proton, in order)
\beqa\label{CIBval}
\Delta a_{\rm CIB} &=& \frac{1}{2} \left( a_{nn} + a_{pp} \right) - a_{np}
= 5.7 \pm 0.3~{\rm fm}~,\nonumber\\
\Delta r_{\rm CIB} &=& \frac{1}{2} \left( r_{nn} + r_{pp} \right) - r_{np}
= 0.05 \pm 0.08~{\rm fm}~.
\eeqa
These numbers for charge independence breaking (CIB)
are based on the Nijmegen potential and the Coulomb effect
for $pp$ scattering is subtracted based on standard methods. 
The charge independence breaking in the scattering lengths is large, of
the order of 25\%, since $a_{np} = (-23.714 \pm 0.013)\,$fm. In addition, there are
charge symmetry breaking (CSB) effects leading to different values for
the $pp$ and $nn$ threshold parameters,
\beqa\label{CSBval}
\Delta a_{\rm CSB} &=&  a_{pp} - a_{nn} 
= 1.5 \pm 0.5~{\rm fm}~,\nonumber\\
\Delta r_{\rm CSB} &=&  r_{pp} - r_{nn} 
= 0.10 \pm 0.12~{\rm fm}~.
\eeqa
Both the CIB and CSB effects have been studied intensively within potential
models of the nucleon--nucleon (NN) interactions. In such approaches, the
dominant CIB comes from the charged to neutral pion mass difference in the
one--pion exchange (OPE), $\Delta a_{\rm CIB}^{\rm OPE} \sim 3.6 \pm 0.2\,$fm. 
Additional
contributions come from $\gamma\pi$ and $2\pi$ (TPE) exchanges. Note also that the
charge dependence in the pion--nucleon coupling constants in OPE and TPE
almost entirely cancel. In the meson--exchange picture, CSB originates mostly from  
$\rho-\omega$ mixing, $\Delta a_{\rm CSB}^{\rho-\omega} \sim 1.2 \pm 0.4\,$fm. Other
contributions due to $\pi-\eta$, $\pi-\eta'$ mixing or the proton--neutron
mass difference are known to be much smaller.

Within QCD, CSB and CIB are of course due to the different masses and charges
of the up and down quarks. Such isospin violating effects can be systematically
analyzed within the framework of chiral effective field theories. In the 
two--nucleon sector, a complication arises due to the unnaturally large
S--wave scattering lengths. This can be dealt with in two ways. One is
the ``hybrid'' approach due to Weinberg~\cite{weinpid}, in which chiral perturbation
theory is applied to the interaction kernel sewed to realistic nuclear
wave functions obtained by conventional means. This approach has 
been successfully applied to a variety
of processes which are dominated by OPE, a particularly striking one being
the prediction of the electric dipole amplitude for neutral pion production
off deuterons~\cite{bblmv} recently measured at SAL~\cite{SAL}. A different and
more {\it systematic} fashion to deal with the unnaturally large scattering
lengths is  the recently proposed power divergence subtraction scheme (PDS)
proposed by Kaplan, Savage and Wise 
(KSW)~\cite{KSW}.\footnote{There exist by now modifications
of this approach and it as been argued that it is equivalent to cut--off
schemes. We do not want to enter this discussion here but rather stick
to its original version.} Essentially, one resums the lowest
order local four--nucleon contact terms $\sim C_0 (N^\dagger N)^2$ (in
the S--waves)
to generate the large scattering lengths and treats the remaining
effects perturbatively, in particular also pion exchange. This means that
most low--energy observables are dominated by contact interactions. 
The chiral expansion for NN scattering entails a new scale $\Lambda_{NN}$ of the
order of 300~MeV, so that one can systematically treat external
momenta up to the size of the pion mass. There have been suggestions
that the radius of convergence can be somewhat enlarged~\cite{mestew},
but in any case $\Lambda_{NN}$ is considerably smaller than the typical
scale of about 1~GeV appearing in the pion--nucleon sector. A status
report of the various observables calculated within this framework can
be found in ref.\cite{KaBonn}.  In this
context, it appears to be particularly interesting to study CIB 
(or in general isospin violation) which is believed to be dominated by
long range pion effects. That is done here.\footnote{For a first
look at these effects in an EFT framework, see the work of van 
Kolck~\cite{birat}\cite{birai}. Electromagnetic corrections to the 
one--pion exchange potential have been considered in~\cite{birar}.}
First, we write down the leading 
strong and electromagnetic four--nucleon contact terms. It is important
to note that in contrast to the pion or pion--nucleon sector, one can 
not easily lump the expansion in small momenta and the electromagnetic coupling
into one expansion but rather has to treat them separately. Then we
consider in detail CIB. The leading effect starts out at order $\alpha
Q^{-2}$, where $Q$ is the generic expansion parameter in the KSW
approach. It stems from OPE plus a contact term of order $\alpha$ with
a coefficient of natural size that scales as $Q^{-2}$. Similarly, the
leading CSB effect are  four--nucleon contact terms of order $\alpha$
and order $m_u -m_d$,
which also scale as $Q^{-2}$. While in the case of $E_0^{(1)}$ this
scaling property is enforced by a cancellation of a divergence, the
situation is a priori different for CSB. However, for a consistent
counting of all isospin breaking effects related to strong or em
insertions, one should count the quark mass difference and virtual
photon effects similarly. Note, however, that these CIB and CSB terms are
numerically much smaller than the leading strong contributions which
scale as $Q^{-1}$ because $\alpha \ll 1$ and $(m_u-m_d)/\Lambda_\chi
\ll 1$.
The corresponding constants, which we call $E_0^{(1,2)}$, together with  the 
strong parameters (as given in the work of KSW) can be determined
by fitting the three scattering lengths
$a_{pp}$, $a_{nn}$, $a_{np}$ and the  $np$ effective range.\footnote{Whenever
we talk of the $pp$ system, we assume that the Coulomb effects have been
subtracted.}
That allows to predict the
momentum dependence of the  $np$ and the $nn$ $^1S_0$ phase shifts.
Based on these observation, we can in addition give a general
classification for the relevant operators contributing to CIB and CSB
in this scheme. Additional work related to long--range Coulomb photon exchange
is necessary in the proton--proton system. We do not deal with this
issue here but refer to recent work using EFT approaches in 
refs.\cite{RK}\cite{BRH}.

\medskip

\noindent {\bf 2.} First, we discuss the various parts of the effective
Lagrangian underlying the analysis of isospin violation in the two--nucleon
system. To include virtual photons in the pion and the pion--nucleon
system is by now a standard procedure~\cite{egpdr}-\cite{mm}.
The lowest order (dimension two)  pion Lagrangian
takes the form
\beq
{\cal L}_{\pi\pi} = \frac{f_\pi^2}{4} \langle \nabla_\mu U \nabla^\mu U^\dagger
+ \chi U^\dagger + \chi^\dagger U \rangle + C \langle QUQU^\dagger\rangle~,
\eeq 
with $f_\pi = 92.4\,$MeV the pion decay constant, $\nabla_\mu$ the (pion) covariant
derivative containing the virtual photons, $\langle \,\, \rangle$ denotes
the trace in flavor space, $\chi$ contains the
light quark masses and the last term, which contains the
nucleon charge matrix $Q$=$e\,$diag$(1,0)$,\footnote{Or equivalently, one can use the
quark charge matrix $e(1+\tau_3)/2$.} leads to the charged to neutral
pion mass difference, $\delta m^2 = m_{\pi^\pm}^2 - m_{\pi^0}^2$, via
$\delta m^2 = 8\pi\alpha C/f_\pi^2$, i.e. $C = 5.9\cdot 10^{-5}\,$GeV$^4$. Note that
to this order the quark mass difference $m_u - m_d$ does not appear in the
meson Lagrangian (due to G--parity). That is chiefly the reason why the
pion mass difference is almost entirely an electromagnetic (em) effect.
The equivalent pion--nucleon Lagrangian to second order takes the form
\beqa
{\cal L}_{\pi N}^{\rm str} &=& N^\dagger \left( iD_0 - \frac{g_A}{2} 
\vec{\sigma}\cdot\vec{u} \right) N +
 N^\dagger \left\{ \frac{\vec{D}^2}{2M} + c_1 \langle \chi_+ \rangle
+\left( c_2 -\frac{g_A^2}{8M}\right) u_0^2 + c_3 u_\mu u^\mu
\right. \nonumber\\
&& \qquad \qquad + \left.\frac{1}{4} 
\left(c_4 + \frac{1}{4M}\right) [\sigma_i , \sigma_j ] u_i u_j
+ c_5 \left( \chi_+ - \frac{1}{2} \langle \chi_+ \rangle\right) + \dots
\right\}  N~,
\eeqa
which is the standard heavy baryon effective Lagrangian in the rest--frame
$v_\mu =(1,0,0,0)$. $M$ is the nucleon mass and $u_\mu$ the chiral
viel--bein, $u_\mu \sim -i\partial_\mu\phi/f_\pi + \ldots\,$. 
The four--nucleon interactions to be discussed below
do not modify the form of this Lagrangian (for a general discussion, see
e.g. ref\cite{FMS}). Strong isospin breaking is due to the operator $\sim c_5$.
Electromagnetic terms to second order are given by~\cite{ms}
\beq\label{LpiN2em}
{\cal L}_{\pi N}^{\rm em} = f_\pi^2 N^\dagger \left\{ f_1 \langle Q_+^2 
- Q_-^2\rangle + f_2 \hat{Q}_+\langle Q_+\rangle + f_3 \langle Q_+^2
+ Q_-^2\rangle + f_4 \langle Q_+\rangle^2 \right\} N~,
\eeq 
with $Q_\pm = uQ^\dagger u \pm u^\dagger Q u^\dagger$ and $\hat{A} = A -
\langle A\rangle /2$ projects onto the off--diagonal elements of the
operator $A$. Evidently, the charge matrices always have to appear
quadratic since a virtual photon can never leave a diagram.  The last two terms
in eq.(\ref{LpiN2em}) are not observable since they lead to an equal em mass
shift for the proton and the neutron, whereas the operator $\sim f_2$ to this 
order gives the em proton--neutron mass difference. In what follows, we will
refrain from writing down such types of operators which only lead
to an overall shift of masses or coupling constants. We note that in the pion
and pion--nucleon sector, one can effectively count the electric charge as
a small momentum or meson mass. This is based on the observation that
$M_\pi / \Lambda_\chi \sim e/\sqrt{4\pi} = \sqrt{\alpha} \sim 1/10$ 
since $\Lambda_\chi
\simeq 4\pi f_\pi = 1.2\,$GeV. It is thus possible to turn the dual
expansion in small momenta/meson masses on one side and in the electric coupling
$e$ on the other side into an expansion with one generic small parameter.
We also remark that from here we use the fine structure constant $\alpha
=e^2/4\pi$ as the em expansion parameter.

We now turn to the two--nucleon sector, i.e. the four--fermion contact
interactions without pion fields. Consider first the strong terms. 
Up to one derivative, the effective Lagrangian takes the form 
\beqa
{\cal L}_{NN}^{\rm str}&=& l_1 (N^\dagger N)^2 + l_2 (N^\dagger \vec{\sigma} N)^2
+ l_3 ( N^\dagger \langle \chi_+ \rangle N) (N^\dagger N) + l_4
(N^\dagger \hat{\chi}_+ N)(N^\dagger N)
\nonumber \\
&+& l_5 ( N^\dagger \vec{\sigma}\langle \chi_+ \rangle N) (N^\dagger \vec{\sigma}
 N) + l_6 (N^\dagger \vec{\sigma} \hat{\chi}_+ N)(N^\dagger \vec{\sigma} N)
+ \ldots~,
\eeqa
where the ellipsis denotes terms with two (or more) derivatives acting on the
nucleon fields. Similarly, one can construct the em terms. The ones without
derivatives on the nucleon fields  read
\beqa
{\cal L}_{NN}^{\rm em}&=& N^\dagger \left\{ r_1 \langle Q_+^2 - Q_-^2\rangle
+ r_2 \hat{Q}_+ \langle Q_+\rangle \right\} N (N^\dagger N) 
\nonumber \\
&+& N^\dagger \vec{\sigma} \left\{ r_3 \langle Q_+^2 - Q_-^2\rangle 
+ r_4  \hat{Q}_+ \langle Q_+\rangle \right\} N (N^\dagger \vec{\sigma} N)
\nonumber \\
&+&   N^\dagger  \left\{ r_{5}Q_+ + r_{6}\langle Q_+ \rangle \right\} 
N (N^\dagger Q_+ N) +
   N^\dagger \vec{\sigma} \left\{ r_{7}Q_+ + r_{8}\langle Q_+ \rangle 
\right\} N (N^\dagger \vec{\sigma} Q_+ N) \nonumber \\
&+& r_{9} (N^\dagger  Q_+ N)^2 + r_{10} (N^\dagger \vec{\sigma} Q_+ N)^2~.
\eeqa
There are also various terms resulting form the insertion of
the Pauli isospin matrices $\vec{\tau}$ in different ${N}^\dagger N$
binomials. Some of these can be eliminated by Fierz reordering, while
the others are of no importance for our considerations.
Note that from now on we consider em effects. The leading CSB $\sim
m_u-m_d$ has the same structure as the corresponding em term and thus
its contribution can be effectively absorbed in the value of
$E_0^{(2)}$, as defined below.
We remark since $\Lambda_{NN}$ is significantly smaller than $\Lambda_\chi$,
it does not pay to treat the expansion in the generic KSW momentum
$Q$ simultaneously with the one in the fine structure constant (as it is done
e.g. in the pion--nucleon sector). Instead, one has
to assign to each term a double expansion parameter $Q^n \alpha^{m}$, with
$n$ and $m$ integers.
Lowest order charge independence breaking is due to a term $\sim (N^\dagger
\tau^3 N)^2$ whereas charge symmetry breaking at that order is given
by a structure $\sim (N^\dagger \tau^3 N) (N^\dagger N)$.
In the KSW approach, it is customary to project the Lagrangian terms on the
pertient NN partial waves. Denoting by $\beta$ the  $^1S_0$ partial
wave for a given cms energy $E_{\rm cms}$, the Born amplitudes for the lowest
order CIB and CSB operators between the various two--nucleon states
takes the form
\beqa\label{LNNME}
\langle \beta , pp | {\cal L}^{\rm em}_{NN} | \beta , pp\rangle &=&
-\left(\frac{Mp}{2\pi}\right) \alpha \left( E_0^{(1)} +   E_0^{(2)} \right)~,
\nonumber\\
\langle \beta , nn | {\cal L}^{\rm em}_{NN} | \beta , nn\rangle &=&
-\left(\frac{Mp}{2\pi}\right) \alpha \left( E_0^{(1)} -   E_0^{(2)} \right)~,
\eeqa
where we will determine the coupling constant $E_0^{(1,2)}$ later on and also derive the
scaling properties of the $E_{2n}^{(1,2)}$. The  terms with the superscript '$(1)$'
refer to CIB whereas the second ones
relevant for CSB  are denoted by the superscript '$(2)$'. Higher order operators
are denoted accordingly. There is, of course, also a CIB contribution to the
$np$ matrix element. To be consistent with the charge symmetric calculation
of ref.\cite{KSW}, we absorb its effect in the constant $D_2$, i.e. it
amounts to a finite renormalization of $D_2$ and is thus not observable.
In eq.(\ref{LNNME}), $p= \sqrt{ME_{\rm cms}}$ is the nucleon cms momentum.

\medskip

\noindent {\bf 3.} Consider now the effect of the charged to neutral
pion mass difference $\delta m^2$, see the upper left diagramm
in fig.~1. OPE between two neutrons or two
protons can obviously involve charged and neutral pions. The mass
difference can be treated in two ways. As proposed in
ref.\cite{mms}, one can modify the pion propagator,
\beq \label{prop}
\Delta_\pi^{ab} (\ell)
= {i \delta^{ab}\over [\ell^2 - m_{\pi^0}^2 
- \delta m^2  \, ( 1- \delta^{3a} \, )]} \,\, ,
\quad \delta m^2 = {8\pi \alpha C \over f_\pi^2} \,\, ,
\eeq
with $\ell$ the pion four--momentum  and $'a,b\,'$ isospin indices.
Since we are interested only in the leading corrections $\sim \delta
m^2 \sim \alpha$, it suffices to work with the expanded form of
eq.(\ref{prop}),
\beq\label{propex}
\Delta_\pi^{ab} (\ell) = {i \delta^{ab}\over \ell^2 - m_{\pi^0}^2}
+ \delta m^2 \frac{i \delta^{ab} ( 1- \delta^{3a} )}{(\ell^2 -
  m_{\pi^0}^2)^2} + {\cal O}\left((\delta m^2)^2\right)~.
\eeq
{}From this we conclude that OPE diagrams with different pion masses
have the isospin structure $O_{12} = \tau_{(1)a} \Delta_\pi^{ab} \tau_{(1)b}
= (a+b)\, \vec{\tau}_{(1)} \,  \vec{\tau}_{(2)}
-b \, {\tau}_{(1)}^3 \, {\tau}_{(2)}^3$ and lead to CIB since 
\beqa
\langle pp | O_{12} | pp \rangle_{\rm Cs} &=&
\langle nn | O_{12} | nn \rangle =
 \frac{a}{4}~, \nonumber\\
\langle np  | O_{12} | np \rangle  &+& \langle np  | O_{12} | pn \rangle 
= \frac{a}{4}+\frac{b}{2}~, 
\eeqa
for the various isospin components of the two--nucleon
system and 'Cs' stands for Coulomb--subtracted. Obviously, 
these effects are of order $\alpha Q^{-2}$. The
$np$ amplitude was already calculated by KSW. We have worked out the
leading corrections $\Delta {\cal A} = {\cal A}_{nn} - {\cal A}_{np}
= {\cal A}_{pp}^{\rm Cs} - {\cal A}_{np}$
due to the pion mass difference. The pertinent diagrams are shown
in fig.~1. We follow the notation of KSW in that we call these
corresponding three amplitudes ${\cal A}_{1,-2}^{II,III,IV}$ where
the first (second) subscript refers to the power in $\alpha$ ($Q$) and
the superscripts to the first three diagrams of the figure. We
find
\beqa
\Delta {\cal A}^{II}_{1,-2} &=&  \Gamma \left[ \frac{1}{4p^2} \ln
\left(1 + \frac{4p^2}{m^2}\right) - \frac{1}{m^2+4p^2} \right]~,
\nonumber\\
\Delta {\cal A}^{III}_{1,-2} &=& \Gamma \left( \frac{mM{\cal
      A}_{-1}}{4\pi}\right) \left[ \frac{1}{pm} \arctan
  \frac{2p}{m} + \frac{i}{2pm}\ln \left( 1 + + \frac{4p^2}{m^2}\right)
  - \frac{1 + \frac{2ip}{m}}{m^2+4p^2} 
\right]\nonumber\\
\Delta {\cal A}^{IV}_{1,-2} &=& \Gamma \left( \frac{mM{\cal
      A}_{-1}}{4\pi}\right)^2 \left[ \frac{i}{m^2} \arctan
  \frac{2p}{m} - \frac{1}{2m^2}\ln \left( \frac{m^2+4p^2}{\mu^2}\right)
  +\frac{1}{m^2} -\frac{1}{2} \frac{1 + \frac{2ip}{m}}{m^2+4p^2} 
\right]\nonumber\\
\Gamma &=& -\delta m^2 \frac{g_A^2}{2f_\pi^2}~, \quad \delta m^2 =
m_{\pi^\pm}^2 - m_{\pi^0}^2~, m^2 =  m_{\pi^0}^2 + 2\delta m^2~,
\eeqa
with ${\cal A}_{-1} \equiv {\cal A}_{0,-1}$ the 
leading term in the expansion of the $np$ $^1S_0$ amplitude~\cite{KSW}
\beq
{\cal A}_{-1} = - \frac{C_0 (\mu )}{1+C_0 (\mu) M (\mu + ip)/4\pi}~.
\eeq
Here,  $\mu$ is the PDS
regularization scale. Note while the diagrams II and III are
convergent, IV diverges logarithmically. Therefore, the Lagrangian
must contain a counterterm of the structure 
$E_0^{(1)} (\mu ) \alpha (N^\dagger \tau^3 N)(N^\dagger \tau^3 N)$
(cf. section~2) since it is needed to make the amplitude
scale--independent. Note that for graph IV we have used the
same subtraction as performed in~\cite{KSW}.
Consequently, for operators of this type with $2n$ derivatives we
can establish the scaling property $E_{2n}^{(1)} \sim Q^{-2+n}$. This does
not contradict the KSW power counting for the isospin symmetric theory
since $\alpha \ll 1$. Stated differently, the leading CIB term
of order $\alpha Q^{-2}$ is numerically much smaller than the
strong leading order contribution $\sim Q^{-1}$. 
The insertion from this contact
term is shown in the last diagram of fig.~1 and leads to an
additional contribution to $\Delta {\cal A}$. In complete analogy,
we can treat the leading order CSB effect which is due to an
operator of the form $\alpha E_0^{(2)}(N^\dagger \tau_3 N)(N^\dagger N)$.
This term is, however, finite. Putting pieces together, we get 
\beq\label{AE}
\Delta {\cal A}^{VI}_{1,-2,{\rm pp}} = -\alpha \left(E_0^{(1)} +E_0^{(2)}\right)
\left[ \frac{{\cal A}_{-1}}{C_0}\right]^2~, \quad
\Delta {\cal A}^{VI}_{1,-2,{\rm nn}} = -\alpha \left(E_0^{(1)} -E_0^{(2)}\right)
\left[ \frac{{\cal A}_{-1}}{C_0}\right]^2~, \quad
\eeq
where the coupling constants $E_0^{(1,2)} (\mu )$ obeys the renormalization group
equations,
\beqa
\mu \, \alpha \,\frac{dE_0^{(1)} (\mu )}{d\mu} 
&=& \alpha \frac{M}{2\pi} E_0^{(1)} (\mu ) 
C_0 (\mu )\mu - \delta m^2
\frac{M^2 g_A^2}{32 \pi^2 f_\pi^2} C_0^2 (\mu)~, \nonumber\\   
\mu \, \alpha \, \frac{dE_0^{(2)} (\mu )}{d\mu} &=& 
\alpha \frac{M}{2\pi} E_0^{(2)} (\mu )  C_0(\mu ) \mu~.
\eeqa
Note that from here on we do no longer exhibit the scale dependence of the
various couplings constants $E_0^{(1,2)}, C_{0,2}, D_2$.
We can now  relate the $pp$ and $nn$ scattering lengths to the $np$ one
(of course, in the $pp$ system Coulomb subtraction is assumed),
\beqa
\frac{1}{a_{pp}} &=&  \frac{1}{a_{np}}  -\frac{4\pi\alpha (E_0^{(1)}
+ E_0^{(2)})}{MC_0^2} + \Delta~,\nonumber\\
\frac{1}{a_{nn}} &=&  \frac{1}{a_{np}}  -\frac{4\pi\alpha (E_0^{(1)}
- E_0^{(2)})}{MC_0^2} + \Delta~,\\
\Delta &=& \delta m^2 \frac{g_A^2 (-C_0 M(m-2\mu ) + 8\pi + C_0 M m \ln
  (m^2/\mu^2))}{16 \pi m C_0 f_\pi^2}~.\nonumber
\eeqa
For the effective ranges, we have only CIB  
\beq
r_{nn} = r_{pp} =  r_{np} + 
\delta m^2 \frac{g_A^2 (C_0 M \mu  +4\pi )( C_0 M (3\mu -2 m) 
+ 12\pi)}{6 \pi M m^4 C_0^2 f_\pi^2}~.
\eeq
Note that this last relation is scale--independent and that it does not contain
any unknown parameter. We remark that for the CIB scattering lengths difference
the pion contribution alone is {\it not} scale--independent and can thus
never be uniquely disentangled from the contact term contribution $\sim E_0^{(1)}$.
While the leading OPE contribution resembles the result obtained in meson 
exchange models, the mandatory appearance of this contact term is a distinctively 
new feature of the effective field theory approach. 
It is easy to classify the leading and next--to--leading em corrections
to these results. At order $\alpha Q^{-1}$, one has  the contribution
from two potential pions with
the  pion mass difference and also contact interactions with two derivatives.
Effects due to the charge dependence of the pion--nucleon coupling constants,
i.e. isospin breaking terms from ${\cal L}_{\pi N}^{\rm em}$, only start to
contribute at order $\alpha Q^0$. Such effects are therefore suppressed by two
orders of $Q$ compared to the leading terms. This finding is in agreement
with the various numerical  analyses performed in potential models.
We now turn to CSB. Here, to leading order there is simply a four--nucleon
contact term proportional to the constant $E_0^{(2)}$.
Its value can be determined from a fit to the empirical
value given in eq.(\ref{CSBval}). First corrections to the leading
order CSB effect are classified below.

\medskip

\noindent {\bf 4.} We now turn to the numerical analysis considering 
exclusively the $^1S_0$ channel. At leading order,
all parameters can be fixed from the pertinent scattering length.
Already in ref.\cite{KSW} it was pointed out that there are various 
possibilities of fixing the next--to--leading order constants. One is
to stay at low energies (i.e. fitting $a$ and $r$) or performing an
overall fit to the phase shift up to momenta of about 200~MeV. In this
latter case, however, the resulting low energy parameters deviate from
their empirical values and also, at $p \sim 150$~MeV, the correction
is as big as the leading term. Since we are interested in small effects
like CIB and CSB, we stick to the former approach and use the   $^1S_0$ 
$np$ phase shift around $p = 0$ to fix the parameters 
as shown by the dashed curve in fig.2. More precisely, we fit the
parameters $C_{0,2}, D_2$ to the $np$ phase shift under the condition
that the scattering length and effective range are exactly reproduced.
The two new parameters $E_0^{(1,2)}$ are determined from the $nn$ and
$pp$ scattering lengths. The resulting parameters at $\mu = m$ are of natural size,
\beqa\label{Cval}
C_0 &=& -3.46~{\rm fm}^2~, \quad
C_2 =  2.75~{\rm fm}^4~, \quad
D_2 =  0.07~{\rm fm}^4~, \nonumber \\
E_0^{(1)} &=&  -6.47~{\rm fm}^2~, \quad
E_0^{(2)} =   1.10~{\rm fm}^2~.
\eeqa
To arrive at the curves shown in fig.2, we have
used the physical masses for the proton and the neutron. We stress
again that the effect of the em terms $\sim E_0^{(1,2)}$ is small
because of the explicit factor of $\alpha$ not shown in eq.(\ref{Cval}). 
Having fixed these parameters, we can now predict the $nn$ $^1S_0$ phase 
shift as depicted by the solid line in fig.2. It agrees with $nn$
phase shift extracted from the Argonne V18 potential (with the
scattering length and effective range exactly reproduced) up to
momenta of about 100~MeV. The analogous curve for
the $pp$ system (not shown in the figure) is close to the solid line
since the CSB effects are very small.

\medskip

\noindent {\bf 5.}
Finally, we can give a summary of the various leading (LO), next--to--leading
(NLO) and next--to--next--to--leading order (NNLO) contributions to
CIB and CSB, with respect to the expansion in  $Q$, to leading order
in $\alpha$ and the light quark mass difference. Consider first CIB. 
\begin{description}
\item[${\rm LO}$, $\alpha Q^{-2}$]Electromagnetic corrections to one--pion exchange
(pion mass difference) and electromagnetic four--nucleon contact interactions
with no derivatives. 
\item[${\rm NLO}$, $\alpha Q^{-1}, \varepsilon Q^{-1}$]
Em corrections to two--pion exchange,
four--nucleon contact terms with two derivatives and insertions
proportional to the strong neutron--proton mass splitting.
\item[${\rm NNLO}$, $\alpha Q^{0}$]Em corrections to three--pion exchange and 
four--nucleon contact terms with four derivatives as well as charge dependent 
coupling constants from the electromagnetic pion--nucleon Lagrangian.
\end{description}
Note that there are no CIB effects due to the light quark mass difference
linear in $\varepsilon = m_u - m_d$,\footnote{One could also use the dimensionsless
small parameter $\epsilon = (m_u-m_d)/(m_u+m_d) \simeq 1/3$, see e.g. 
ref.\cite{birat}.
We prefer to keep the notation which is most commonly
used in the pion--nucleon sector, see e.g. ref.\cite{bkmrev}.}
 except the trivial effects related to the strong $pn$
mass difference. 
We now turn to CSB. The pattern of the various contributions looks different:
\begin{description}
\item[${\rm LO}$, $\alpha Q^{-2}, \varepsilon Q^{-2}$]Electromagnetic
  and strong isospin--breaking 
four--nucleon contact interactions with no derivatives. 
\item[${\rm NLO}$, $\alpha Q^{-1}, \varepsilon Q^{-1}$]
Em four--nucleon contact terms with two derivatives, strong isospin--breaking
contact interaction with two derivatives and insertions proportional to the
neutron--proton mass difference.
\end{description}
Here, we assume the same scaling in powers of $Q$ for the em and the
strong coupling constants.
We remark that the leading order CSB effects do not modify the effective
range but the corresponding scattering length.

\medskip

\noindent {\bf 6.}
In summary, we have considered electromagnetic and strong isospin violation
in low--energy nucleon--nucleon scattering in the effective field theory
formalism developed in ref.\cite{KSW}. In particular, the leading charge
independence breaking effect is due to a combination of the neutral to charged
pion mass difference in one--pion exchange diagrams together with an electromagnetic
four--nucleon contact term. Its corresponding coupling constant scales as $Q^{-2}$
but is numerically suppressed by the explicit appearance of the fine structure
constant $\alpha \sim 1/137$. We have shown how the KSW power counting has to
be modified in the presence of isospin violating operators.
We explicitely evaluated the $^1S_0$ phase shifts for the $np$, $nn$ and 
Coulomb--subtracted $pp$ systems at next--to--leading order. 
In addition, we have given a general
classification of the various CIB and CSB corrections. It would be interesting
to extend this formalism to other partial waves and to higher energies
so as to investigate e.g. isospin violation in pion production.
\bigskip

\section*{Acknowledgments}

\noindent We thank Vincent Stoks for providing us with the $nn$ phase
shift.

\bigskip



\vspace{2cm}

\section*{Figures}

\vspace{2cm}

\begin{figure}[htb]
\centerline{\epsfig{file=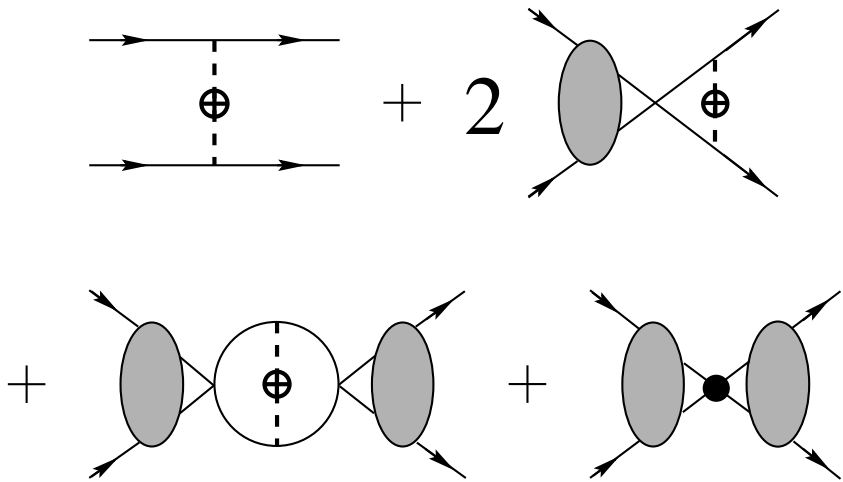,width=2.5in,angle=270}}

\vspace{3.3cm}

\caption[Adiag]{\protect \small
Relevant graphs contributing to charge independence breaking at 
leading order $\alpha Q^{-2}$. The blob stands for the resummation
of the lowest order $(N^\dagger N)^2$ contact terms $\sim C_0$. The open (filled)
circle denotes a pion mass insertion $\sim \delta m^2$ (an insertion
of the leading four--nucleon operators $\sim \alpha E_0^{(1)}$). For
charge symmetry breaking, the leading contribution is given by the
last diagram with the filled circle denoting an insertion$\sim \alpha E_0^{(2)}$.}
\end{figure}

\pagebreak

$\,$

\vspace{5cm}

\begin{figure}[htb]
\centerline{\epsfig{file=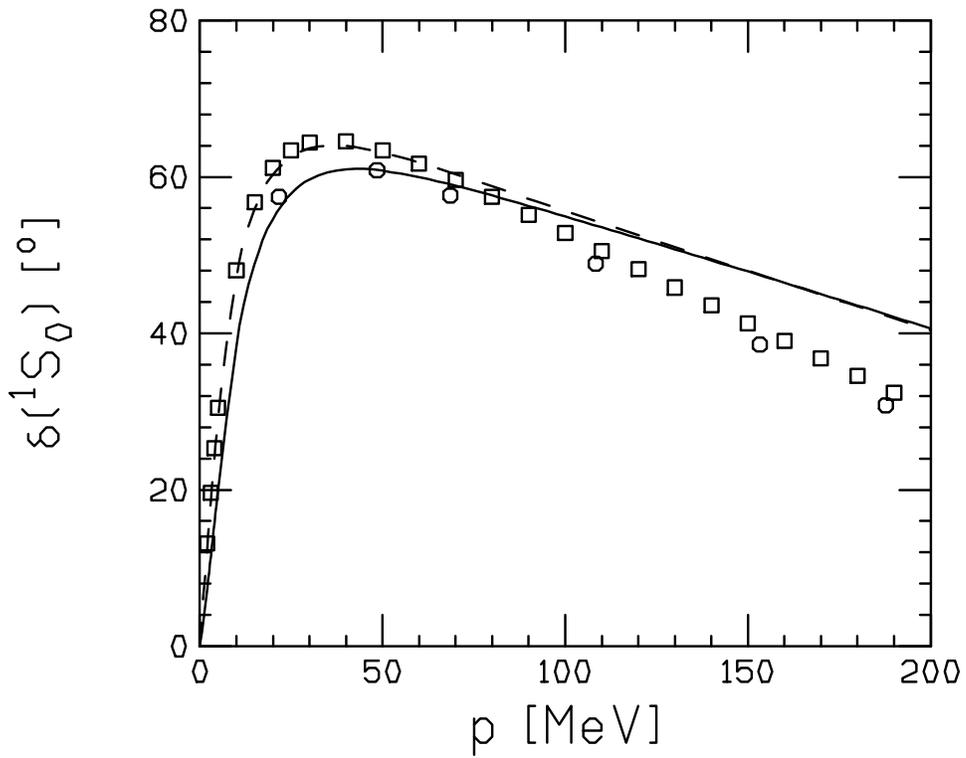,width=5in,}}

\vspace{2.3cm}

\caption[delnnp]{\protect \small
$^1S_0$ phase shifts for the $np$ (dashed line) and $nn$ (solid
line) systems versus the nucleon cms momentum. 
The emipirical values for the $np$ case (open squares)
are taken from the Nijmegen analysis~\cite{nij}. The open octagons
are the $nn$ ``data'' based on the Argonne V18 potential.}
\end{figure}

\end{document}